\documentstyle[12pt]{article}

\catcode`@=11
\newcount\@tempcntc
\def\@citex[#1]#2{\if@filesw\immediate\write\@auxout{\string\citation{#2}}\fi
  \@tempcnta\z@\@tempcntb\m@ne\def\@citea{}\@cite{\@for\@citeb:=#2\do
    {\@ifundefined
       {b@\@citeb}{\@citeo\@tempcntb\m@ne\@citea\def\@citea{,}{\bf ?}\@warning
       {Citation `\@citeb' on page \thepage \space undefined}}%
    {\setbox\z@\hbox{\global\@tempcntc0\csname b@\@citeb\endcsname\relax}%
     \ifnum\@tempcntc=\z@ \@citeo\@tempcntb\m@ne
       \@citea\def\@citea{,}\hbox{\csname b@\@citeb\endcsname}%
     \else
      \advance\@tempcntb\@ne
      \ifnum\@tempcntb=\@tempcntc
      \else\advance\@tempcntb\m@ne\@citeo
      \@tempcnta\@tempcntc\@tempcntb\@tempcntc\fi\fi}}\@citeo}{#1}}
\def\@citeo{\ifnum\@tempcnta>\@tempcntb\else\@citea\def\@citea{,}%
  \ifnum\@tempcnta=\@tempcntb\the\@tempcnta\else
   {\advance\@tempcnta\@ne\ifnum\@tempcnta=\@tempcntb \else \def\@citea{--}\fi
    \advance\@tempcnta\m@ne\the\@tempcnta\@citea\the\@tempcntb}\fi\fi}
\catcode`@=12

\begin{document}

\title{\vskip-3cm{\baselineskip14pt
\centerline{\normalsize DESY 00-123\hfill ISSN 0418-9833}
\centerline{\normalsize hep-ph/0010159\hfill}
\centerline{\normalsize September 2000\hfill}}
\vskip1.5cm
Order $\alpha^7\ln(1/\alpha)$ Contribution to Positronium Hyperfine Splitting} 
\author{{\sc Bernd A. Kniehl} and {\sc Alexander A. Penin}\thanks{Permanent
address: Institute for Nuclear Research, Russian Academy of Sciences,
60th October Anniversary Prospect 7a, Moscow 117312, Russia.}\\
{\normalsize II. Institut f\"ur Theoretische Physik, Universit\"at Hamburg,}\\
{\normalsize Luruper Chaussee 149, 22761 Hamburg, Germany}}

\date{}

\maketitle

\thispagestyle{empty}

\begin{abstract}
The logarithmically enhanced $\alpha^7\ln(1/\alpha)$ contribution to the
hyperfine splitting of the positronium ground-state energy levels is
calculated in the framework of dimensionally regularized nonrelativistic
quantum electrodynamics. 
The correction is negative and amounts to about 1/3 of the leading logarithmic
$\alpha^7\ln^2(1/\alpha)$ one.
The discrepancy between the experimental measurements and the theoretical
prediction is reduced.
\medskip

\end{abstract}

\newpage

Positronium, which is an electromagnetic bound state of the electron $e^-$ and
the positron $e^+$, is the lightest known atom.
Thanks to the smallness of the electron mass $m_e$, strong- and
weak-interaction effects are negligible, and its properties can be calculated
perturbatively in quantum electrodynamics (QED), as an expansion in
Sommerfeld's fine-structure constant $\alpha$, with very high precision, only
limited by the complexity of the calculations.
Positronium is thus a unique laboratory for testing the QED theory of weakly
bound systems.
However, the theoretical analysis is complicated due to annihilation and
recoil effects.

The positronium hyperfine splitting (HFS)
$\Delta\nu=E\left(1^3S_1\right)-E\left(1^1S_0\right)$,
where $E\left(1^1S_0\right)$ and $E\left(1^3S_1\right)$ are the energy levels
of para- and orthopositronium, respectively, is the most precisely measured
quantity in positronium spectroscopy as far as the absolute precision is
concerned.
The most recent measurements of the HFS \cite{Mil,Rit} yielded
\begin{eqnarray}
\Delta\nu^{\rm exp}&=&203.387\,5(16)\,\mbox{GHz},
\label{exp1}\\
\Delta\nu^{\rm exp}&=&203.389\,10(74)\,\mbox{GHz},
\label{exp2}
\end{eqnarray}
respectively.

The present theoretical knowledge may be summarized as:
\begin{eqnarray}
\Delta\nu^{\rm th}&=&\Delta\nu^{\rm th}_0\left\{1
-\frac{\alpha}{\pi}\left(\frac{32}{21}+\frac{6}{7}\ln2\right)
+\frac{5}{14}\alpha^2\ln\frac{1}{\alpha}\right.
\nonumber\\
&&{}+\left(\frac{\alpha}{\pi}\right)^2
\left[\frac{1367}{378}-\frac{5197}{2016}\pi^2
+\left(\frac{6}{7}+\frac{221}{84}\pi^2\right)\ln2\right.
\nonumber\\
&&{}-\left.\frac{159}{56}\zeta(3)\right]
-\frac{3}{2}\,\frac{\alpha^3}{\pi}\ln^2\frac{1}{\alpha}
+C\frac{\alpha^3}{\pi}\ln\frac{1}{\alpha}
\nonumber\\
&&{}+\left.D\left(\frac{\alpha}{\pi}\right)^3\right\},
\label{theo}
\end{eqnarray}
where $\Delta\nu^{\rm th}_0=7m_e\alpha^4/12$ is the leading-order result
\cite{Pir}.
The first-order correction was calculated in Ref.~\cite{KarKle}.
The logarithmically enhanced $\alpha^6\ln(1/\alpha)$ term was found in
Ref.~\cite{BY,CasLep1}.  
The nonlogarithmic ${\cal O}(\alpha^6)$ term includes the contribution due to
the radiative correction to the Breit potential \cite{BroEri}, the three-,
two- and one-photon annihilation contributions \cite{ABZ}, the
non-annihilation radiative recoil contribution \cite{STY}, and the pure recoil
correction computed numerically in Ref.~\cite{Pac1} and analytically in
Ref.~\cite{CMY}.
In ${\cal O}(\alpha^7)$, only the leading double-logarithmic
$\alpha^7\ln^2(1/\alpha)$ term is available \cite{Kar}.

Including all the terms known so far, we have
\begin{equation}
\Delta\nu^{\rm th}=203.392\,01\,\mbox{GHz},
\label{ini}
\end{equation}
which exceeds Eqs.~(\ref{exp1}) and (\ref{exp2}) by approximately 2.8 and 3.9
experimental standard deviations, respectively.
In contrast to the well known {\it orthopositronium lifetime 
puzzle},\footnote{For the most recent developments of this problem, see, for
example, Ref.~\cite{KniPen1} and the references cited therein.}
the experimental situation for the HFS is unambiguous.
In fact, the experimental error is compatible with a naive estimate of the
theoretical uncertainty due to as-yet unknown higher-order corrections.
Should this discrepancy persist after the dominant terms of the latter have
been calculated, this would provide a signal for {\it new physics}. 
This makes the HFS to be one of the most interesting topics in positronium
spectroscopy, both from the experimental and theoretical points of view.

Thus, it is an urgent matter to improve the prediction of the HFS as much as
possible, and one is faced with the task of analyzing the third-order
correction, which is extremely difficult.
However, there is a special subclass of the ${\cal O}(\alpha^7)$ contributions
which can be analyzed separately, namely those which are enhanced by powers of
$\ln(1/\alpha)\approx5$. 
They may reasonably be expected to provide an essential part of the full
${\cal O}(\alpha^7)$ contributions.
This may be substantiated by considering Eq.~(\ref{theo}) in
${\cal O}(\alpha^6)$, where the logarithmic term is approximately 2.6 times
larger than the nonlogarithmic one.
While the leading double-logarithmic ${\cal O}(\alpha^7)$ contribution to
Eq.~(\ref{theo}) is known \cite{Kar}, the subleading single-logarithmic one is
yet to be found.
In fact, from the positronium lifetime calculation \cite{KniPen1,HilLep,MelYel}
we know that the subleading terms can be as important as the leading ones.
The purpose of this Letter is complete our knowledge of the logarithmically
enhanced terms of ${\cal O}(\alpha^7)$ by providing the coefficient $C$ in
analytic form.

The origin of the logarithmic corrections is the presence of several scales
in the bound-state problem.
The dynamics of the nonrelativistic $e^+e^-$ pair near threshold involves
four different scales:
(i) the hard scale (energy and momentum scale like $m_e$);
(ii) the soft scale (energy and momentum scale like $\beta m_e$);
(iii) the potential scale (energy scales like $\beta^2m_e$, while 
momentum scales like $\beta m_e$); and
(iv) the ultrasoft scale (energy and momentum scale like $\beta^2m_e$).
Here $\beta$ denotes the electron velocity in the center-of-mass frame.
The logarithmic integration over a loop momentum between different scales 
yields a power of $\ln(1/\beta)$.
Since positronium is approximately a Coulomb system, we have 
$\beta\propto\alpha$. 
This explains the appearance of powers of $\ln(1/\alpha)$ in 
Eq.~(\ref{theo}). 
The leading logarithmic corrections may be obtained straightforwardly by
identifying the regions of logarithmic integration \cite{CasLep1,Kar}.
The calculation of the subleading logarithms is much more involved because
certain loop integrations must be performed exactly beyond the logarithmic
accuracy. 

In the following, we briefly outline the main features of our method
developed previously in Ref.~\cite{KniPen1}, where it was applied to the
analysis of the subleading logarithmic third-order corrections to the
positronium ground-state decay rates.
This approach is similar to the one adopted in Ref.~\cite{MelYel}. 
We work in nonrelativistic QED (NRQED) \cite{CasLep2}, which is the effective
field theory that emerges by expanding the QED Lagrangian in $\beta$ and
integrating out the hard modes.
If we also integrate out the soft modes and the potential photons, we arrive
at the effective theory of potential NRQED \cite{PinSot1}, which contains
potential electrons and ultrasoft photons as active particles.
Thus, the dynamics of the nonrelativistic $e^+e^-$ pair is governed by the
effective Schr\"odinger equation and by its multipole interaction with the
ultrasoft photons.
The corrections from harder scales are contained in the higher-dimensional
operators of the nonrelativistic Hamiltonian, corresponding to an expansion in
$\beta$, and in the Wilson coefficients, which are expanded in $\alpha$. 
In the process of scale separation, spurious infrared and ultraviolet
divergences arise, which endow the operators in the nonrelativistic
Hamiltonian with anomalous dimensions.
In fact, these divergences completely determine the logarithmic corrections
\cite{KniPen1,KniPen3}.
We use dimensional regularization, with $d=4-2\epsilon$ space-time dimensions,
to handle these divergences, which are of the form $1/\epsilon^n$
($n=1,2,\ldots$) as $\epsilon\to0$ \cite{CMY,PinSot2,BenSmi}.
Compared to the {\it traditional} NRQED approach, endowed with an explicit
momentum cutoff and a fictitious photon mass to regulate the ultraviolet and
infrared behavior \cite{CasLep1,HilLep,NioKin1}, this scheme has the advantage
that contributions from different scales are matched automatically. 
 
In the effective theory, the HFS is generated by spin-flip operators of the
effective nonrelativistic Hamiltonian averaged over the bound-state wave
function.
The hard-scale corrections, which require fully relativistic QED calculations
and are most difficult to find, do not depend on $\beta$ and do not lead to
logarithmic contributions by themselves.
However, they can interfere with the logarithmic corrections from the softer
scales.
The only results from relativistic perturbation theory that enter our analysis
are the one-loop renormalizations of the relevant operators in the effective
nonrelativistic Hamiltonian. 
The missing ingredients can all be obtained in the nonrelativistic
approximation.
The leading-order ${\cal O}(\alpha^4)$ HFS is generated by the
${\cal O}(\beta^2)$ spin-flip part of the tree-level Breit potential, which
consists of the Fermi operator $V_F$ and the annihilation operator
$V_{\rm ann}$. 
The $\alpha^7\ln(1/\alpha)$ contribution, which we are interested in, arises
from several sources.
A part of it can be extracted from the positronium lifetime calculation
\cite{KniPen1}.
This part corresponds to: (1) the second-order corrections in nonrelativistic
Rayleigh-Schr\"odinger perturbation theory, which arise from (1a) the
insertions of the tree-level ${\cal O}(\beta^2)$ spin-independent Breit
potential and the one-loop hard corrections to the leading-order spin-flip
operators and (1b) the insertions of the one-loop ${\cal O}(\alpha\beta^2)$
operators related to the hard \cite{MPS}, soft \cite{PinSot2}, or ultrasoft
\cite{KniPen1,KniPen2} scales (cf.\ Eqs.~(13)--(15) of \cite{KniPen1},
respectively) and the leading-order spin-flip operators; 
(2) {\it irreducible} corrections to the leading-order spin-flip operators,
which include (2a) ${\cal O}(\alpha\beta^2)$ and (2b)
${\cal O}(\alpha^2\beta)$ terms and can be obtained from the corresponding
equations of Ref.~\cite{KniPen1} by replacing the leading-order decay operator
$V_4({\bf p},{\bf p^\prime},{\bf S})$ by $V_F+V_{\rm ann}$.
The hard renormalization coefficients of the leading-order operators $V_F$
and $V_{\rm ann}$ are 1 and $-(44/9+2\ln2)$, respectively
\cite{NioKin1,LZB}.
They replace the renormalization coefficients $A_{p,o}$ of
$V_4({\bf p},{\bf p^\prime},{\bf S})$ in Ref.~\cite{KniPen1}.
Another nontrivial difference with respect to the positronium lifetime
calculation \cite{KniPen1} is that the Fermi operator can cause a $\Delta L=2$
transition of the spin-triplet state, so that $D$-wave intermediate states
contribute to second order in nonrelativistic perturbation theory, by double
insertion of $V_F$.
This gives an additional contribution of 5/42 to the coefficient $C$.
Finally, there is a modification of the soft part of contribution (1b) due to
the spin-dependent part of the transverse-photon-exchange contribution, which
vanishes for the annihilation channel. 
This gives an additional contribution of 68/63 to the coefficient $C$. 
Using the results of Ref.~\cite{KniPen1} and taking these modifications into
account, we obtain the following contribution to the coefficient $C$ from the
sources enumerated above
\begin{eqnarray}
C_I&=&
\left[-\frac{64}{21}-\frac{12}{7}\ln{2}\right]_{1a}
+\left[-\frac{8}{15}+\frac{8}{3}\ln2\right]_{1b}
+\left[\frac{4}{3}\right]_{2a}
\nonumber\\
&&{}+\left[\frac{41}{6}-\frac{32}{3}\ln2\right]_{2b}
+\left[\frac{5}{42}\right]_{D-{\rm wave}}+
\left[\frac{68}{63}\right]_{\rm soft}
\nonumber\\
&=&\frac{1822}{315}-\frac{68}{7}\ln2,
\label{part1}
\end{eqnarray}  
where the various contributions are given separately.
Note that contribution (1b) also includes the entire double-logarithmic term
not presented in Eq.~(\ref{part1}).
The structure of the overlapping divergences resulting in the
double-logarithmic contribution to the HFS is similar to the positronium
lifetime analysis \cite{KniPen1}.

Another part of the $\alpha^7\ln(1/\alpha)$ contribution is produced in the
first order of the nonrelativistic Rayleigh-Schr\"odinger perturbation theory
by the one-loop hard corrections to the ${\cal O}(\beta^4)$ and
${\cal O}(\alpha\beta^3)$ spin-flip operators, which give rise to the
$\alpha^6\ln(1/\alpha)$ contribution to the HFS.
The relevant operators are generated by the relativistic correction to the
Coulomb-photon exchange, the relativistic correction to the transverse-photon
exchange, the kinematical retardation, and the one-loop correction involving
{\it seagull} vertex diagrams with one Coulomb and one transverse photon.
For our calculation, we only need the three hard renormalization coefficients,
usually denoted as $c_F$, $c_S$, and $c_{pp^\prime}$, which are related to the
anomalous magnetic moment of the electron and are finite and scheme
independent.
By gauge invariance, the  coefficients $c_F$ and $c_S$ are the same as for the
${\cal O}(\beta^2)$ operators, while the coefficient $c_{pp^\prime}$
parameterizes the new ${\cal O}(\alpha\beta^4)$ operator contribution to the
transverse-photon exchange \cite{NioKin1}.
The resulting contributions to the coefficient $C$ are
\begin{eqnarray}
C_{II}&=&\left[\frac{1}{14}\right]_{\rm Cou}
+\left[-\frac{5}{14}\right]_{\rm tra}
+\left[\frac{2}{7}\right]_{\rm ret}
+\left[-\frac{12}{7}\right]_{\rm sea}
\nonumber\\
&=&-1.
\end{eqnarray}
The remaining part of the $\alpha^7\ln(1/\alpha)$ contribution also
corresponds to the first order of the nonrelativistic Rayleigh-Schr\"odinger
perturbation theory and is related to the relativistic corrections to the
operators contributing to the HFS in lower orders.
The relevant operators are generated by the diagrams with one or two
transverse-photon exchanges, where the momentum of the photon with
spin-independent interaction can be soft or ultrasoft, and by the soft
diagrams with one or two seagull vertices involving either one Coulomb and one
transverse photon or two transverse photons. 
The resulting contribution to the coefficient $C$ reads
\begin{eqnarray}
C_{III}&=&-\frac{41}{63}.
\label{part3}
\end{eqnarray}

Adding Eqs.~(\ref{part1})--(\ref{part3}), we obtain 
\begin{eqnarray}
C&=&\frac{62}{15}-\frac{68}{7}\ln2\approx-2.6001.
\end{eqnarray}
Thus, the $\alpha^7\ln(1/\alpha)$ term in Eq.~(\ref{theo}) has the same sign 
as the $\alpha^7\ln^2(1/\alpha)$ one and amounts to about 1/3 of the latter.
It reduces $\Delta\nu$ by 323~kHz, while the $\alpha^7\ln^2(1/\alpha)$ term
reduces $\Delta\nu$ by 918~kHz.
For comparison, we recall that, in the counterpart of Eq.~(\ref{theo}) 
appropriate for the muonium HFS, the coefficient of the
$\alpha^7\ln^2(1/\alpha)$ correction reads $-8/3$ and
$C=281/180-(8/3)\ln2\approx-0.2873$ \cite{LayZwa}.
Our final prediction for the HFS reads
\begin{equation}
\Delta\nu^{\rm th}=203.391\,69(41)\,\mbox{GHz}.
\label{fin}
\end{equation}
Here, the uncertainty due to the unknown nonlogarithmic ${\cal O}(\alpha^7)$ 
term in Eq.~(\ref{theo}) is estimated by using the value $D=16.233\pi^2$ of 
the analogous coefficient in the case of the HFS of muonium
\cite{NioKin1,Pac2}.

The unknown nonlogarithmic ${\cal O}(\alpha^7)$ term in Eq.~(\ref{theo})
receives contributions from three-loop QED diagrams with a considerable number
of external lines, which are still beyond the reach of presently available
computational techniques.
In this sense, we expect Eq.~(\ref{fin}) to remain the best prediction for the
foreseeable future.

The new theoretical value in Eq.~(\ref{fin}) exceeds the experimental values 
in Eqs.~(\ref{exp1}) and (\ref{exp2}) by approximately 2.6 and 3.5
experimental standard deviations, respectively.
Thus, the discrepancy between experiment and theory is somewhat reduced by the
inclusion of the $\alpha^7\ln(1/\alpha)$ term, but is still remains sizeable.

We may speculate about the magnitude of the coefficient $D$ in
Eq.~(\ref{theo}).
Note that two powers of $\alpha$ in the nonlogarithmic ${\cal O}(\alpha^7)$
term can be of nonrelativistic origin.
Each of them should be accompanied by the characteristic factor $\pi$, which
happens for the logarithmic terms.
Thus, a plausible estimate of the coefficient $D$ is a few units times 
$\pi^2$.
In order to bring the theoretical estimate into agreement with
Eqs.~(\ref{exp1}) and (\ref{exp2}), we need $D\approx-100\pi^2$ and
$D\approx-70\pi^2$, respectively.
On the other hand, the logarithmic terms of the positronium HFS exhibit a 
structure similar to the muonium case, so that it is not unreasonable to 
expect the nonlogarithmic terms of the positronium and muonium HFS's to be of
the same magnitude.
This would imply a significant contradiction between the current experimental
measurements and the theoretical prediction.
However, although this may seem unlikely, one cannot completely exclude the
possibility that the residual discrepancy will be removed by the inclusion of
the nonlogarithmic ${\cal O}(\alpha^7)$ term. 
We conclude that both the calculation of the coefficient $D$ as well as
improved experimental measurements are necessary in order to establish or
remove the residual discrepancy.
Although there is no conceptual problem on the theoretical side, from the
technical point of view, such a calculation represents a challenge for QED
bound-state perturbation theory.   

Finally we would like to note that the technique developed in
Ref.~\cite{KniPen1} and here can be applied to the analysis of QCD heavy
quark-antiquark bound states, where the logarithmically enhanced corrections
are known to be essential \cite{KniPen3,BPSV}.

We are indebted to K. Melnikov and A. Yelkhovsky for useful discussions.
This work was supported in part by the Deutsche Forschungsgemeinschaft through
Grant No.\ KN~365/1-1, by the Bundesministerium f\"ur Bildung und Forschung
through Grant No.\ 05~HT9GUA~3, and by the European Commission through the
Research Training Network {\it Quantum Chromodynamics and the Deep Structure
of Elementary Particles} under Contract No.\ ERBFMRX-CT98-0194.
The work of A.A.P. was supported in part by the Volkswagen Foundation through
Grant No.\ I/73611.


\begin{thebibliography}{99}

\bibitem{Mil}
A. P. Mills, Jr.\ and G. H. Bearman, 
Phys.\ Rev.\ Lett.\ {\bf34}, 246 (1975);
A. P. Mills, Jr.,
Phys.\ Rev.\ A {\bf27}, 262 (1983).

\bibitem{Rit}
M. W. Ritter, P. O. Egan, V. W. Hughes, and K. A. Woodle,
Phys.\ Rev.\ A {\bf30}, 1331 (1984).

\bibitem{Pir} 
J. Pirenne, 
Arch.\ Sci.\ Phys.\ Nat.\ {\bf29}, 265 (1947);
V. B. Berestetski and L. D. Landau,    
Zh.\ Eksp.\ Teor.\ Fiz.\ (USSR) {\bf19}, 673 (1949);
R. A. Ferrell, 
Phys.\ Rev.\ {\bf84}, 858 (1951).

\bibitem{KarKle}
R. Karplus and A. Klein, 
Phys.\ Rev.\ {\bf87}, 848 (1952).

\bibitem{BY} 
G. T. Bodwin and D. R. Yennie, 
Phys.\ Rep.\ {\bf43}, 267 (1978).

\bibitem{CasLep1} 
W. E. Caswell and G. P. Lepage, 
Phys.\ Rev.\ A {\bf20}, 36 (1979).

\bibitem{BroEri}
S. J. Brodsky and G. W. Erickson, 
Phys.\ Rev.\ {\bf148}, 26 (1966);
R. Barbieri, J. A. Mignaco, and E. Remiddi, 
Nuovo Cim.\ {\bf11A}, 824 (1972).

\bibitem{ABZ}
G. S. Adkins, M. H. T. Bui, and D. Zhu,
Phys.\ Rev.\ A {\bf37}, 4071 (1988).
G. S. Adkins, Y. M. Aksu, and M. H. T. Bui,
Phys.\ Rev.\ A {\bf47}, 2640 (1993);
G. S. Adkins, R. N. Fell, and P. M. Mitrikov,
Phys.\ Rev.\ Lett.\ {\bf79}, 3383 (1997);
A. H. Hoang, P. Labelle, and S. M. Zebarjad,
Phys.\ Rev.\ Lett.\ {\bf79}, 3387 (1997).

\bibitem{STY}
J. R. Sapirstein, E. A. Terray, and D. R. Yennie,
Phys.\ Rev.\ D {\bf29}, 2290 (1984);
K. Pachucki and S. G. Karshenboim,
Phys.\ Rev.\ Lett.\ {\bf80}, 2101 (1998).

\bibitem{Pac1} 
K. Pachucki, 
Phys.\ Rev.\ A {\bf56}, 297 (1997);
G. S. Adkins and J. Sapirstein, 
Phys.\ Rev.\ A {\bf58}, 3552 (1998); Erratum {\it ibid.}, to be published;
A. P. Burichenko, 
Report No.\ hep-ph/0004063.

\bibitem{CMY}
A. Czarnecki, K. Melnikov, and A. Yelkhovsky,
Phys.\ Rev.\ Lett.\ {\bf82}, 311 (1999);
Phys.\ Rev.\ A {\bf59}, 4316 (1999).

\bibitem{Kar}
S. G. Karshenboim,
Zh.\ Eksp.\ Teor.\ Fiz.\ {\bf103}, 1105 (1993)
[Sov.\ Phys.\ JETP {\bf76}, 541 (1993)].

\bibitem{KniPen1}
B. A. Kniehl and A. A. Penin,
Phys.\ Rev.\ Lett.\ {\bf85}, 1210 (2000); {\bf85}, 3065(E) (2000).

\bibitem{HilLep}
R. Hill and G. P. Lepage,
Report No.\ hep-ph/0003277 v2 (May 2000).

\bibitem{MelYel}
K. Melnikov and A. Yelkhovsky,
Report No.\ SLAC-PUB-8557 and hep-ph/0008099.

\bibitem{CasLep2}
W. E. Caswell and G. P. Lepage,
Phys.\ Lett.\ {\bf167B}, 437 (1986).

\bibitem{PinSot1}
A. Pineda and J. Soto,
Nucl.\ Phys.\ Proc.\ Suppl.\ {\bf64}, 428 (1998).

\bibitem{KniPen3}
B. A. Kniehl and A. A. Penin,
Nucl.\ Phys.\ {\bf B577}, 197 (2000).

\bibitem{PinSot2}
A. Pineda and J. Soto,
Phys.\ Lett.\ B {\bf420}, 391 (1998);
Phys.\ Rev.\ D {\bf59}, 016005 (1999).

\bibitem{BenSmi}
M. Beneke and V. A. Smirnov,
Nucl.\ Phys.\ {\bf B522}, 321 (1998).

\bibitem{NioKin1}
T. Kinoshita and M. Nio,
Phys.\ Rev.\ D {\bf53}, 4909 (1996).

\bibitem{MPS}
A. V. Manohar,
Phys.\ Rev.\ D {\bf56}, 230 (1997);
A. Pineda and J. Soto,
Phys.\ Rev.\ D {\bf58}, 114011 (1998).

\bibitem{KniPen2}
B. A. Kniehl and A. A. Penin,
Nucl.\ Phys.\ {\bf B563}, 200 (1999).

\bibitem{LZB} 
P. Labelle, S. M. Zebarjad, and  C. P. Burgess,  
Phys.\ Rev.\ D {\bf56}, 8053 (1997).

\bibitem{LayZwa} 
D. E. Zwanziger,
Bull.\ Am.\ Phys.\ Soc.\ {\bf6}, 514 (1961);
Nuovo Cim.\ {\bf XXXIV}, 77 (1964);
A. J. Layzer,
Bull.\ Am.\ Phys.\ Soc.\ {\bf6}, 514 (1961);
Nuovo Cim.\ {\bf XXXIII}, 1538 (1964).

\bibitem{Pac2}
T. Kinoshita and M. Nio,
Phys.\ Rev.\ Lett.\ {\bf72}, 3803 (1994);
M. I. Eides and V. A. Shelyuto,  
Pis'ma Zh.\ Eksp.\ Teor.\ Fiz.\ {\bf61}, 465 (1995)
[JETP Lett.\ {\bf61}, 478 (1995)];
Phys.\ Rev.\ A {\bf52}, 954 (1995);
K. Pachucki,
Phys.\ Rev.\ A {\bf54}, 1994 (1996);
M. Nio and T. Kinoshita,
Phys.\ Rev.\ D {\bf55}, 7267 (1997).

\bibitem{BPSV}
N. Brambilla, A. Pineda, J. Soto, and A. Vairo,
Phys.\ Lett.\ B {\bf470}, 215 (1999).

\end{thebibliography}
\end{document}